\documentclass[aps,preprint,amsmath,amssymb,amsfonts,nofootinbib]{revtex4-1}
\usepackage{epsfig}
\usepackage{graphicx}
\usepackage{dcolumn}
\usepackage{bm}
\usepackage{amsthm}
\usepackage{amsmath}
\usepackage{color}
\newcommand{\beq}{\begin{equation}}
\newcommand{\eeq}{\end{equation}}
\newcommand{\bea}{\begin{eqnarray}}
\newcommand{\eea}{\end{eqnarray}}

\begin{document}
\title{The Anomalous Scaling Exponents of Turbulence in General Dimension from Random Geometry}
\author{Christopher Eling}
\email{cteling@gmail.com}
\affiliation{Rudolf Peierls Centre for Theoretical Physics, University of Oxford, 1 Keble Road, Oxford OX1 3NP, UK}

\author{Yaron Oz}
\email{yaronoz@post.tau.ac.il}
\affiliation{Raymond and Beverly Sackler School of Physics and Astronomy, Tel Aviv University, Tel Aviv 69978, Israel}

\date{\today}
\begin{abstract}
We propose an analytical formula for the anomalous scaling exponents of inertial range
structure functions in incompressible
fluid turbulence. The formula is a Knizhnik-Polyakov-Zamolodchikov (KPZ)-type relation and is valid in any number of space dimensions.
It incorporates intermittency in a novel way by dressing the Kolmogorov linear scaling via a coupling to a lognormal random geometry.
The formula has one real parameter $\gamma$ that depends on the number of space dimensions. The scaling exponents satisfy the convexity
inequality, and the supersonic bound constraint.
They agree with the experimental
and numerical data in two and three space dimensions, and with numerical data in four space dimensions.  Intermittency
increases with $\gamma$, and in the infinite $\gamma$ limit
the scaling exponents approach the value one, as in Burgers turbulence.
At large $n$ the $n$th order exponent scales as $\sqrt{n}$. We discuss the relation between fluid flows and black hole geometry that inspired our proposal.

\end{abstract}

\pacs{...}

\maketitle

The remarkable phenomenon of fluid turbulence is one of the major unsolved problems of physics  \cite{Frisch}.
Most fluid motions
in nature at all scales are turbulent.
Aircraft motions, river flows, atmospheric phenomena, astrophysical flows and even blood flows
are some examples of set-ups where turbulent flows occur.
Despite centuries of research, we still lack
an analytical description and understanding of fluid flows in the non-linear regime.
Insights into turbulence hold a key to understanding the principles and
dynamics of  non-linear systems with a large number of strongly interacting degrees of freedom far from equilibrium.
In addition to being a major challenge  to basic science, understanding turbulence is likely to
have an important impact on diverse practical problems ranging from environmental issues such as pollution and concentration of chemicals to cardiovascular
physiology.

In this paper we will mainly consider incompressible fluid flows in $d\geq 2$ space dimensions. They are the relevant flows when the velocities are much smaller
than the speed of sound.
The incompressible Navier-Stokes (NS) equations provide a mathematical formulation of the fluid flow evolution. They read
\begin{equation}
\partial_t v^i + v^j\partial_j v^i = -\partial_i p + \nu \partial_{jj} v^i,~~~~~~\partial_iv^i = 0,~~~ i=1,...,d  \ ,
\label{NS}
\end{equation}
where $v^i$  is the fluid velocity and  $p$ is the fluid pressure. An important dimensionless parameter in the study of fluid flows  is the Reynolds number ${\cal R}_e = \frac{l v}{\nu}$, where $l$
is a characteristic length scale, $v$ is the velocity difference at that scale, and $\nu$ is the kinematic viscosity.
The Reynolds number quantifies the relative strength of the
non-linear interaction compared to the viscous term.
When the Reynolds number is of order a thousand or more,
one observes numerically and experimentally a turbulent structure of the flow.
This phenomenological observation
is general, and fluid details are of no importance.
The turbulent velocity field exhibits highly complex spatial and temporal structures and  appears to be a random process. Thus, even
though the NS equations are deterministic (in the absence of a random force),
a single realization of a solution to the NS equations is unpredictable.

Instead of studying individual solutions to the NS equations, one is led to consider the statistics of the solutions.
The statistics can be defined in various ways. One can use an ensemble average by averaging over initial conditions. Turbulence that is reached
in this way is a decaying one. Alternatively, one can introduce
a random force. This allows reaching a sustained steady state turbulence with an energy source
and a viscous sink.
The statistical properties of turbulent flows are remarkable.
Numerical and experimental data show that the statistical average properties
exhibit a universal structure shared by all turbulent flows, independently
of the details of the flow excitations.
One defines the inertial range to be the range of distance scales $L_V\ll r \ll L_F$,  where the scales $L_V$ and $L_F$ are
determined by the viscosity and forcing, respectively.
Turbulence at the inertial range of scales  reaches a steady state that exhibits statistical homogeneity and isotropy.

One defines the longitudinal velocity difference between points separated by a fixed distance
$r=|\vec{r}|$
\begin{equation}
\delta v(r) = \left(\vec{v}(\vec{r},t) - \vec{v}(0,t)\right)\cdot \frac{\vec{r}}{r}  \ .
\end{equation}
The structure functions $S_n(r) = \langle (\delta v(r))^n \rangle$
exhibit in the inertial range a scaling $S_n(r) \sim r^{\xi_n}$.

The exponents $\xi_n$  are universal, and depend only on the number of space dimensions.
In 1941 Kolmogorov \cite{Kolmogorov} argued that in three space dimensions the incompressible non-relativistic fluid dynamics in the inertial range follows a cascade breaking of large eddies
to smaller eddies, called a direct cascade,  where energy flux is being transferred from large eddies to small eddies without dissipation.
He further assumed scale invariant
statistics, that is
\begin{align}
P(\delta v(r) )\delta v(r) = F\left(\frac{\delta v(r)}{r^h}\right),
\end{align}
where $P(\delta v(r) )$ is the probability density function, and
$h$ is  a real parameter.
Treating the mean viscous energy dissipation rate $\epsilon$ as a constant in the limit of infinite Reynolds number, he deduced a linear scaling of the exponents $\xi_n = n/3$.

All direct cascades are known numerically and experimentally to break scale invariance and do not simply follow
Kolomogorov scaling. Note, that in two space dimensions the energy cascade is inverse, that is the energy flux is instead transferred to large scales.
Kolmogorov's assumption that the random velocity field is self-similar is incorrect in direct cascades, but it seems to hold in
the inverse cascade.
The self-similarity assumption misses the intermittency of the turbulent flows.
Thus, in order to calculate the scaling exponents one has to quantify the inertial range intermittency effects. The calculation of the anomalous exponents and their deviation from the Kolmogorov scaling is a major open problem.

Since 1941 many multifractal models of turbulence have been proposed \cite{Frisch}. These express multifractality directly in terms of fluctuations of the velocity increments or of the energy dissipation.  For example, Kolmogorov and Obukhov \cite{K62,Obukhov} proposed to replace the constant global average $\epsilon$ with local averages $\epsilon_r$  over a volume of dimension $r$. One then considers $\epsilon_r$ as a lognormally distributed random variable with variance $\sigma^2 \sim -ln(r)$. Later, Mandelbrot \cite{Mandelbrot} argued that one should think about the energy cascade as a random multiplicative process. In this case a random measure can be formalised mathematically as a limit of a ``Gaussian multiplicative cascade" (see \cite{multi}).The lognormal model assumes ``refined self-similarity"
\begin{align}
P(\delta v(r) )\delta v(r) = F\left(\frac{\delta v(r)}{(\langle \epsilon_r \rangle r)^{1/3}}\right),
\end{align}
which leads to a formula for $\xi_n$ producing physically inconsistent supersonic velocities at large $n$ and a violation of the convexity inequality \cite{Frisch}.

Our proposal in this paper is different. We propose that in intermittent turbulence, Kolmogorov linear scaling itself is evaluated with respect to a lognormal random measure.
This is a ``gravitational" dressing of Kolmogorov scaling which is inspired by the relation of fluid dynamics and black hole horizon dynamics
in one higher space dimension \cite{Eling:2010vr}. We propose the dressing of Kolmogorov scaling is via a KPZ (Knizhnik-Polyakov-Zamolodchikov)-type relation \cite{Knizhnik:1988ak}.
This gives an analytical formula for the scaling exponents of incompressible fluid turbulence in any number of space
dimensions $d\geq 2$. It reads
\begin{equation}
\xi_n -\frac{n}{3}= \gamma^2(d) \xi_n(1-\xi_n) \ ,
\label{KPZ}
\end{equation}
where $\gamma(d)$ is a numerical real parameter that depends on the number of space dimensions $d$.

A major part of the paper will be devoted to checking our proposed formula (\ref{KPZ}).
We will first verify that the scaling exponents $\xi_n$ obtained from (\ref{KPZ}) satisfy the convexity
inequality and the supersonic bound constraint.
We will then show that they agree with the experimental
and numerical data in two and three space dimensions, and with the numerical data in four space dimensions.
Intermittency
increases with $\gamma$, and  in the infinite $\gamma$ limit
the scaling exponents approach the value one, as in Burgers turbulence. At large $n$ the $n$th order exponent scales as $\sqrt{n}$.

We will show that the formula does not apply to the  Kraichnan model for passive scalar advection by a random velocity field \cite{passive1,passive2}.
This may have been expected since physics of passive scalar turbulence is different from the incompressible Navier-Stokes system. On the other hand, it is
possible that the more general type of random measure can describe passive scalar turbulence.

The paper is organized as follows.
In section II we will explain the coupling to a random geometry, discuss the proposed formula and its properties, and
perform analytical checks and comparison to experimental and numerical data.
While we will establish certain properties of the function $\gamma(d)$, we will not calculate its precise form in the paper.
In section III we will apply the formula to the passive scalar model.
In section IV we will discuss the relation between fluid flows and black hole geometry that inspired our proposal.
Section IV is devoted to a discussion and open problems.

\section{Exact Formula for the Scaling Exponents}

\subsection{Coupling to a Random Geometry}

Coupling to a random geometry means changing the Euclidean measure $d x$ on a $R^d$ to a random measure $d\mu_{\gamma}(x) = e^{\gamma \phi(x) -\frac{\gamma^2}{2}} d x$, where the Gaussian random field $\phi(x)$ has covariance $\phi(x)\phi(y) \sim - \log|x-y|$ when $|x-y|$ is small (but still in the inertial range). Physically, the notion of distance $r$ is modified in the new measure. Consider a set of scaling exponents (Hausdorff dimensions in the mathematical setup) $\xi_0$ with respect to the  Euclidean measure. Denote the same set of exponents, but now with respect to the random measure, by $\xi$. Then $\xi$ and $\xi_0$ are related by the KPZ relation
\begin{equation}
\xi -\xi_0= \gamma^2(d) \xi(1-\xi).
\label{KPZ1}
\end{equation}
Mathematically, this is a known method to obtain a multifractal structure from a fractal one (for a review see \cite{multi} and
references therein). Our proposal  is that one can incorporate the effect of intermittency at the inertial of range of scales
by coupling to a random geometry in this way and evaluating
the Kolmogorov linear scaling exponents $\xi_0 = \frac{n}{3}$ with respect
to the random measure.

Physically, it is highly nontrivial that the steady state statistics of turbulence can be viewed as such a combination
of the scale invariant statistics and intermittency. Note, that intermittent features appear at short length scales, and this is when the effects of the random field $\phi$ are prominent.
We conjecture that the $e^{\gamma \phi(x)-\frac{\gamma^2}{2}}$ is proportional to local energy flux field $\epsilon(x)$,
\begin{equation}
\epsilon(x) = \frac{\nu}{2} \left(\partial_i v_j +\partial_j v_i\right)^2  \ ,
\end{equation}
in the direct cascade of the turbulent fluid. This has some similarities to the Kolmogorov-Obukhov lognormal model. In that case, refined self-similarity implies the following simple dressing of Kolmogorov scaling
\begin{align}
\langle (\delta v(r))^n \rangle \sim r^{\xi_n} \sim \langle (\epsilon_r)^{n/3} \rangle r^{n/3}.
\end{align}
Evaluating the expectation of the lognormal energy dissipation, one finds
\begin{align}
\xi_n - \frac{n}{3}  = \gamma^2 \frac{n}{3} \left(1-\frac{n}{3} \right). \label{KO}
\end{align}

As noted in \cite{Frisch} this formula fails as it implies $\xi_n$ is a decreasing function for large enough $n$, which violates basic physical inequalities.
The Kolmogorov-Obukhov formula (\ref{KO}) consists of the leading terms in the expansion at small intermittency $\gamma$ of the KPZ formula (\ref{KPZ}).
The two formulas agree up to order $\gamma^2$. The KPZ formula is different for high intermittency and
may be viewed as a completion/generalization of the Kolmogorov-Obukhov formula to the strong intermittency regime\footnote{Note also, that the KPZ formula (\ref{KPZ1}) can be mapped into the lognormal model by exchanging $\xi$ and $\xi_0$ and multiplying by an overall minus sign.}.

Here we assume instead that the fluctuating dissipation field  $\epsilon(x)$, acts as a random measure. Let us make a few comments on the mathematical structure of this coupling to a random geometry. First, note that there are numerical factors that depend on the number of space dimensions, between $\gamma$ appearing
in the random measure and $\gamma$ in (\ref{KPZ1}) \cite{multi}. Since what is relevant for us is the formula (\ref{KPZ1}), we will keep for simplicity
the notation where $\gamma^2$ appears in (\ref{KPZ1}).

The KPZ relation was first derived by coupling a two-dimensional conformal field theory (CFT) to gravity and analyzing the effect
of quantum gravity on the scaling dimensions of the CFT \cite{Knizhnik:1988ak}. This has been dubbed ``gravitational dressing".
The KPZ relation has been generalized in various directions. First, it has been extended to an arbitrary number of dimensions without reference
to a conformal field theory structure \cite{multi}. The KPZ formula is then viewed as a relation between a set of Hausdorff dimensions measured with respect
to a Euclidean (Lebesgue) measure and a random measure.
Second, one can consider a more general random field than the lognormally distributed one \cite{Schramm,Bail}. In this case there is the generalized relation
\begin{align}
\xi_0 = \xi - \log_{2} E[W^{\xi}] ,\label{genKPZ}
\end{align}
where $E$ is the expectation and $W$ is the random variable associated with the measure (not necessarily a lognormal one).
We will not use this generalization in this paper, but it may be valuable
in the study of steady state statistics of other non-linear dynamical systems out of equilibrium.

We will consider the formula (\ref{KPZ1}), where $\gamma$ takes values in the range $[0, \infty)$. However, when $\gamma >1$,
the mathematical construction of the random measure changes. In the two-dimensional quantum gravity language, $\gamma$ is related to the central
charge of the matter system $c$, and the critical value $\gamma=1$
is the $c=1$ barrier. The regime $\gamma >1$ is a different phase of the theory, dubbed a "dual phase".
There may be a duality relation between two phases parametrized by $\gamma$ and $\gamma'$ that
satisfy $\gamma \gamma' = 1$. This could have an interesting impact on the study of turbulence in diverse dimensions.

\subsection{An Exact Formula}

We propose that the scaling exponents of incompressible fluid turbulence $\xi_n$ in any number of space dimensions $d$ satisfy the KPZ-type relation
(\ref{KPZ}). Solving for $\xi_n$ we get
\begin{equation}
\xi_n =  \frac{\left((1+\gamma^2)^2 + 4\gamma^2(\frac{n}{3}-1)\right)^{\frac{1}{2}} + \gamma^2 -1}{2\gamma^2}  \ ,
\label{scalingexp}
\end{equation}
where in choosing the branch we required finite exponents $\xi_n$.
$\gamma(d)$ is a numerical real parameter that depends on the number of space dimensions $d$.  It can be determined from any
moment, for instance, from the energy spectrum.

There are several immediate properties of the formula (\ref{scalingexp})  that we can see.
First, using $n=3$ in (\ref{scalingexp}) one gets the exponent $\xi_3 = 1$ in any dimension, an exact result derived by Kolmogorov which
agrees with numerical simulations and experiments. In \cite{Falkovich:2009mb} this scaling was derived without employing the cascade picture, but simply
from the fact that the NS equations are conservation laws.
Second, the scaling exponent  $\xi_2$ is a monotonically increasing function of $\gamma$, while
the exponents $\xi_n, n>3$ are monotonically decreasing functions of $\gamma$.
Third, in the limit $n\rightarrow 0$ we get that $\xi_n\rightarrow 0$, as expected.
Fourth, in the limit  $\gamma\rightarrow 0$ we have $\xi_n\rightarrow \frac{n}{3}$, that is scale invariant statistics with no intermittency.
Fifth, in the limit $\gamma\rightarrow \infty$, we have $\xi_n\rightarrow 1$, as in Burgers turbulence. The scaling exponents take
values in the range $\frac{2}{3}\leq \xi_2 \leq 1$, and  $1 \leq \xi_n \leq \frac{n}{3}$ for $n\geq 3$. We will propose that the limit $\gamma \rightarrow \infty$,  is the limit of infinite number of space dimensions $d$. The subleading correction, relevant for developing a systematic $\frac{1}{d}$ expansion reads
\begin{equation}
\xi_n = 1 + \frac{1}{\gamma^2}\left(\frac{n}{3}-1\right) + O\left(\frac{1}{\gamma^4}\right)  \ .
\label{sub}
\end{equation}
 Sixth, in the limit $n\rightarrow \infty$ for fixed $\gamma$, we have
 \begin{align}
 \xi_n\rightarrow \frac{1}{\gamma}\left(\frac{n}{3}\right)^{\frac{1}{2}},
 \end{align}
thus growing as $\sqrt{n}$. Seventh, at the "critical" value $\gamma=1$ we get $\xi_n=\left( \frac{n}{3}\right)^{\frac{1}{2}} $.

\subsection{Analytical Constraints on the Scaling Exponents}

 If there exist two consecutive even numbers $2n$ and $2n+2$ such that
$\xi_{2n} > \xi_{2n+2}$, then the velocity of the flow cannot be bounded.
Using (\ref{scalingexp}) it is straightforward to show that $\xi_{2n} \leq \xi_{2n+2}$ for any $\gamma$, thus (\ref{scalingexp}) satisfies the absence of supersonic velocity requirement. The second condition is that of convexity. For any three positive integers $n_1\leq n_2 \leq n_3$, the scaling exponents satisfy the convexity inequality that follows from H\"{o}lder inequality
\begin{align}
(n_3-n_1)\xi_{2n_2} \geq (n_3-n_2)\xi_{2n_1} + (n_2-n_1)\xi_{2n_3}.
\end{align}
Using (\ref{scalingexp}) it is straightforward to show that the H\"{o}lder inequality holds.
Equality is achieved  when $\gamma =0$, when $\gamma \rightarrow \infty$ and when $n_i=n_j$ for some $i\neq j$ and arbitrary $\gamma$.

\subsection{The Energy Spectrum}

The structure function $S_2(r)\sim r^{\xi_2}$ gives the energy spectrum of the fluid.
Using  (\ref{scalingexp}) we see that $\xi_2$ is a monotonic function of $\gamma$ that takes values in the range
$\frac{2}{3} \leq \xi_2 \leq 1$ when $\gamma$ goes from zero to infinity.
In momentum  space a deviation from the Kolmogorov spectrum for small $\gamma$ (small $d$) reads
\begin{equation}
E(k) \sim k^{-\frac{5}{3} - \frac{2\gamma^2}{9}} \ .
\end{equation}
For large $\gamma$  (large $d$) we have
\begin{equation}
E(k) \sim k^{-2 + \frac{1}{3\gamma^2}} \ .
\end{equation}

\subsection{Comparison to Experimental and Numerical Data}

The anomalous scaling exponents  (\ref{scalingexp}) depend on the parameter $\gamma$, which is a function of $d$.
We do not know the exact expression of $\gamma$, but it can be calculated knowing one of the structure functions, such as
the energy spectrum $\gamma = \left(\frac{\xi_2 -\frac{2}{3}}{\xi_2(\xi_2-1)} \right)^{\frac{1}{2}} $.
With this knowledge we can then make an infinite number of predictions. In the following we will compare the analytical expression
(\ref{scalingexp}) to the available numerical and experimental data in various dimensions.

\subsubsection{Two Space Dimensions}

In two space dimensions the energy cascade is an inverse cascade, where the energy flux flows to scales larger than the injection scale.
In this case, one has the energy spectrum agreeing
with the Kolmogorov scaling $\xi_2 = \frac{2}{3}$.
 Using  (\ref{scalingexp}), this implies that $\gamma(2)=0$, and that all the other scaling exponents follow the Kolmogorov scaling $\xi_n = \frac{n}{3}$.

\subsubsection{Three Space Dimensions}

In three space dimensions we first use the data for the anomalous scaling exponents quoted in \cite{Benzi} from wind tunnel experiments at Reynolds number $\sim 10^4$. This experimental data is consistent with numerical data from simulations of the Navier-Stokes equations, see e.g. \cite{Gotoh2002}. We fit  $(\ref{KPZ})$ to this data using a least squares fit with the function FindFit in Mathematica. We see in Figure 1 an excellent agreement, finding that $\gamma^2$ is about 0.161.

\begin{figure}
\begin{center}
 \includegraphics[angle=0,width=12cm]{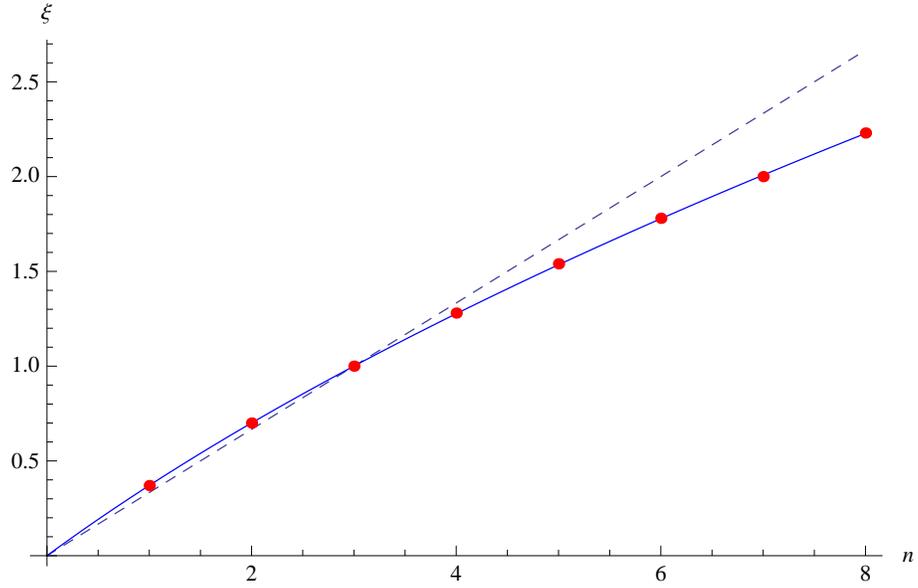}\\
\end{center}
\caption{\label{3dfit}Fit of (\ref{KPZ}) (blue) to experimental data \cite{Benzi} (Table 2). The dashed line represents Kolmogorov scaling. The best fit value of the free parameter $\gamma^2$ is about 0.161. The error on the data is about $\pm 1$ percent.}
\end{figure}

Next, we consider the numerical results for low order structure function exponents and non-integer $n$ given in \cite{Chen}. The numerical data is consistent with experiment at Reynolds number $10^4$. For this data, the fitted value of $\gamma^2$ is about 0.159 and in Figure 2 we again see excellent agreement.
\begin{figure}
\begin{center}
 \includegraphics[angle=0,width=12cm]{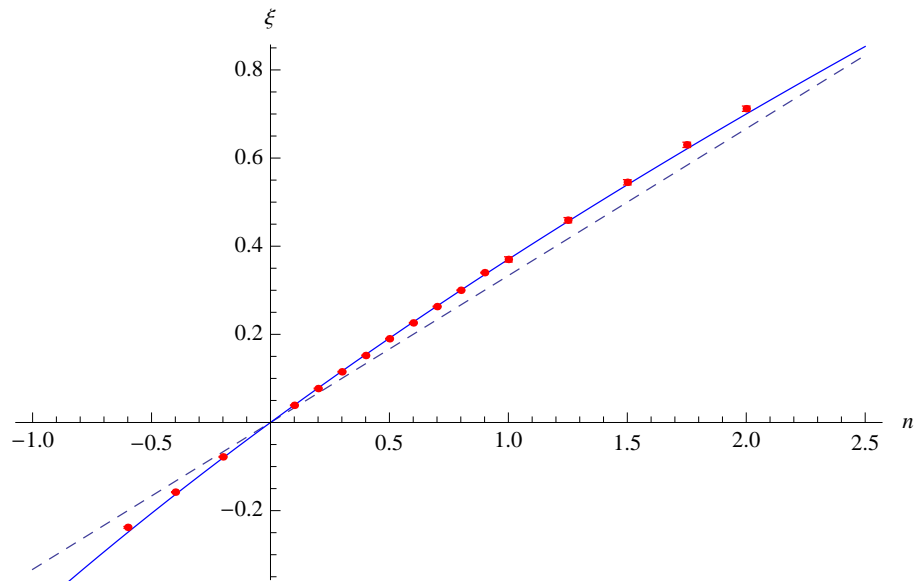}\\
\end{center}
\caption{\label{3dfit}Fit of (\ref{KPZ}) (blue) to numerical data of low moments \cite{Chen} (red). The dashed line represents Kolmogorov scaling. The best fit value of the free parameter $\gamma^2$ is about 0.159.}
\end{figure}
Note that if our conjectured relation between the random measure and the local energy dissipation field is correct, one can determine $\gamma^2$ independently  by measuring the scaling exponent of the two point function, $\langle \epsilon(x)\epsilon(0) \rangle \sim x^{-\gamma^2}$. This value has been found to be $\approx 0.2$, which in our formula is still consistent with the data.

\subsubsection{Four Space Dimensions}
In four space dimensions, numerical simulations of the Navier-Stokes equations were performed in \cite{Gotoh2007}. The authors found an increase in intermittency, i.e. $\xi^{(4)}_{n} > \xi^{(3)}_{n}$ for $n<3$, while $\xi^{(4)}_{n} < \xi^{(3)}_{n}$ for $n>3$. We took the data for the structure function exponents in 4d given in \cite{Gotoh2007} and performed a fit to (\ref{KPZ}). This is shown in Figure 3. Although taken at a relatively low Reynolds number, the results are in agreement with a simple increase in the $\gamma^2$ parameter in our formula (\ref{KPZ}). The value of $\gamma^2$
in four space dimensions is fitted to about 0.278. Note that their numerical data for same simulation in three space dimensions predicts $\gamma^2$ about 0.188, which is higher than the experimental data above. This could be related to the relatively low Reynolds numbers involved.
\begin{figure}
\begin{center}
 \includegraphics[angle=0,width=12cm]{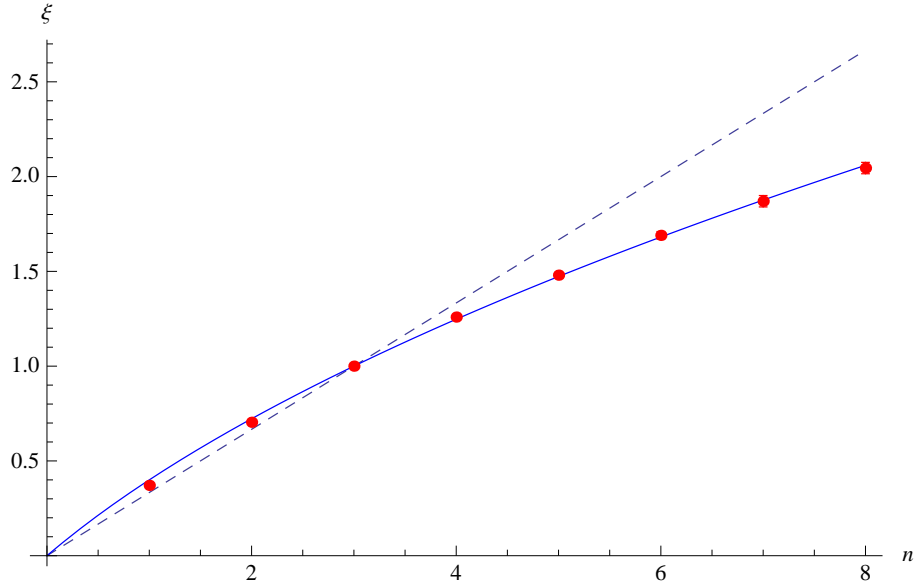}\\
\end{center}
\caption{\label{4dfit.pdf}Fit of (\ref{KPZ}) to the 4d exponents given in \cite{Gotoh2007}. The solid line is the 4d fit with $\gamma^2$ about 0.278.}
\end{figure}

\subsubsection{Intermittency and the Large $d$ Limit}

In order to observe intermittency one has to study the short distance statistical properties of the fluid flow. There are
various measures  for intermittency, such as $F_n(r) = \frac{S_n(r)}{S_2(r)^{\frac{n}{2}}},~~~n \geq 3$.
$F_n(r)$ are expected to grow as a power-law in the limit $r\rightarrow 0$, while staying in the inertial range of scales.

We can analyze the properties of $F_n(r)$ using (\ref{scalingexp}).
They scale as $\sim r^{\alpha}$, where $\alpha$ is a decreasing function of $\gamma$.
In the limit $\gamma \rightarrow 0$ one has $\alpha\rightarrow 0$ and no intermittency, while as
$\gamma \rightarrow \infty$ we get  the maximal intermittency $\alpha = \frac{2-n}{2}$.
Numerically, one sees in \cite{Gotoh2007} a clear growth of $F_n(r), n\geq 4$ in the limit $r\rightarrow 0$, when
 as we increase the number of space dimensions in the simulation. The data is not accurate enough to observe the  growth
 when $n=3$.

 Another exponent that is used to quantify the intermittency is
\begin{equation}
\mu = 2 -\xi_6 \ .
\end{equation}
Experimentally in three space dimensions it has been measured in the range $0.2$ to $0.25$.
Using  (\ref{scalingexp}) with $\gamma^2=0.161$ we get $\mu=0.222$. Expanding around $\gamma=0$ ($d=2$) we have
$\mu = 2 \gamma^2 + o(\gamma^4)$, while expanding around infinite $\gamma$ we have $\mu = 1 - \frac{1}{\gamma^2} + o(\frac{1}{\gamma^4})$.

In \cite{Falkovich:2009mb}, (also see \cite{Gotoh2007}) it was conjectured that in the limit of infinite $d$ all the exponents $\xi_n$ approach the same value, one, as in Burgers turbulence  \cite{Burgers}.
With our formula (\ref{KPZ}) this means that $\gamma$ goes to infinity  in the limit of infinite $d$, and therefore $\xi_n=1$ for any $n$.
This suggests  the interesting possibility of having a systematic  $\frac{1}{d}$ expansion (\ref{sub}).

\section{Passive Scalar Turbulence}

It is natural to ask whether our proposed exact formula for the scaling exponents of incompressible fluid turbulence is applicable
for other systems that exhibit turbulent structure. In the following we will consider
the  Kraichnan model for passive scalar advection by a random Gaussian field of velocities $v^i$, which is
white-in-time  \cite{passive1,passive2}.
The statistics of the velocities is determined by a zero mean $v^i(t,\vec{r}) = 0$, and by the covariance
\begin{equation}
\langle v^i(t,\vec{r})v^i(t,\vec{r}')\rangle = \delta(t-t') D^{ij}(\vec{r},\vec{r}') \ .
\end{equation}
In the inertial range $D^{ij}(\vec{r}) - D^{ij}(0) \sim |\vec{r}|^{\zeta}$, where $\zeta$ takes values between 0 and 2.

Examples of passive scalar systems are smoke in
the air, salinity in the water and temperature when one can neglect thermal
convection.
The evolution equation describes a passively-advected scalar field $T$ driven by the velocity field
\begin{equation}
\partial_t T + v^i\partial_i T = \kappa \partial_{jj}T + f \ ,
\end{equation}
where $\kappa$ is the molecular diffusivity of $T$ and $f$ is an external force.

One defines the dimensionless Peclet number ${\cal P}_e$ as the ratio of the scale of fluctuations of $T$ produced $f$
and the diffusion scale. When  ${\cal P}_e \gg 1$ there is a scalar turbulence with a scalar cascade and
constant flux of $T^2$.
Similarly to the incompressible fluid turbulence, one is interested here in the stationary statistics and the scaling properties of the
scalar structure functions in the inertial range of scales. Define $\delta T(r) = \left(T(\vec{r},t) - T(0,t)\right)$
as the difference between the values of the scalar field at two points separated by a fixed distance
$r=|\vec{r}|$. Then,
\begin{equation}
S_n(r) = \langle \left(\delta T(r)\right)^{2n}\rangle \sim r^{\xi_{2n}} \ .
\end{equation}

Here the scale invariant statistics is Gaussian with $\xi_{2n}=n(2-\zeta)$.
We can now attempt to include the intermittency by the random geometry dressing and the KPZ-type equation
\begin{equation}
\xi_{2n} -n(2-\zeta)= \gamma^2(d) \xi_{2n}(1-\xi_{2n}) \ .
\label{PS}
\end{equation}
Solving for $\xi_{2n}$ we get
\begin{equation}
\xi_{2n} =  \frac{\left((1+\gamma^2)^2 + 4\gamma^2(n(2-\zeta)-1)\right)^{\frac{1}{2}} + \gamma^2 -1}{2\gamma^2}  \ .
\label{scalingps}
\end{equation}

This formula is similar, but not exactly the one proposed by Kraichnan  \cite{passive2}.
In the limit $n\rightarrow \infty$ for fixed $\gamma$, we have
$\xi_{2n}$ growing as $\sqrt{n}$. This is not the expected behaviour, rather $\xi_{2n}$ should approach a constant. This is due to presence of solutions with fronts and shock behaviour, which
seems to be characteristic of compressible fluid systems \cite{passive}. While the KPZ formula does not apply to the passive scalar, it is possible that intermittency in this case ultimately can be described by other, more general types of random geometry, e.g. via the formula (\ref{genKPZ}) above.

\section{Black Hole Horizon Dynamics}

In the following we will briefly review the relation between fluid flows and black hole horizon geometry
 (for a review, see \cite{Eling:2010vr} and references therein),
that inspired our proposal to incorporate the intermittency at the inertial range of scales by a gravitational dressing using a random geometry.
Consider the Einstein equations with a negative cosmological constant $\Lambda$ in $(d+2)$ space-time dimensions

\beq E_{AB} \equiv  R_{AB} - \frac{1}{2} g_{AB} R + \Lambda g_{AB} = 0,~~~~A,B = 0, ...,d+1 \ , \label{Einstein} \eeq

where $g_{AB}$ is the  Lorentzian metric, $R_{AB}$ the Ricci curvature and $R= g^{AB} R_{AB}$ the Ricci scalar.
Black holes are a classical solution of Einstein equations (\ref{Einstein}), and their hallmark is
the existence of a horizon $H$. For example, a black hole solution to the Einstein equations has the form

\begin{align}
ds^2 = g_{AB} dX^A dX^B = -r^2 f(r) dt^2 + 2 dt dr + r^2 \sum^{d}_{i=1} dx_i dx^i \ ,
\end{align}

where the coordinates are $X^A = (t,r,x^i)$. The function $f(r) = 1 - (\frac{4\pi T}{(d+1) r})^{d+1} $.
The horizon is defined as the surface where $f(r)$ vanishes. It is a $(d+1)$-dimensional null hypersurface, forming a causal boundary preventing any light and particles that cross it from returning. Hence, it effectively introduces dissipation.
One can associate with the black hole horizon a temperature $T$ (appearing in $f(r)$), and an entropy proportional to its cross-sectional area $A$.
In Planck units $\hbar=G_N=c=1$, the relation between the area and the entropy is $S_{BH} = \frac{A}{4}$. This structure is called black hole thermodynamics.

Black hole hydrodynamics is a generalization of black hole thermodynamics, similar to the generalization of field theory thermodynamics to hydrodynamics.
While black hole thermodynamics quantifies the thermal equilibrium situation, black hole hydrodynamics describes slow derivations from equilibrium.
In particular, one can allow the black hole temperature to be a slowly varying function $T(t,x^i)=const.(1 + p(t,x^i))$,
and consider black hole itself to be moving at velocity $v^i(t,x)$ with respect to some rest frame.
The perturbed solution to the Einstein equations in this setting will yield a slowly evolving curved geometry,
with the gravity variables providing a geometrical framework for studying the dynamics of fluids.
This can be made precise in the context of a holographic correspondence, where the fluid system lives on a $(d+1)$ dimensional surface of $r=const.$
in the $(d+2)$ dimensional bulk solution.

The motion of fluids translates to the evolution of the black hole horizon, and the fluid
variables to its geometrical data. The normal vector ${\bf n}$ to the horizon $H$ satisfies
$g_{AB} n^A n^B = 0$. The horizon hypersurface is defined by $r = r_H = const.$ and we have

\begin{equation}
n^r =0,~~~~n^t=1,~~~~ n^i = v^i \ . \label{normalform}
\end{equation}

Thus there is a geometrical representation of the fluid velocity in terms of the normal to the black hole horizon hypersurface.
As we perturb the black hole and get it out of equilibrium, the horizon location changes, and up to an overall constant can be parametrized
by $r_H \sim 1+p(t,x)$. The variable $p(t,x)$ quantifies the deviation from equilibrium and is identified as the fluid pressure. The deviation of the horizon area measure from equilibrium is parametrized by the pressure. One gets that the horizon area measure $\sqrt{\gamma}$ scales like $d p(t,x^i)$, where $d$ is the number of space dimensions.

The set of the Einstein equations projected on the horizon
\beq
 E_{AB} n^A = 0 \ ,
\eeq
describe the evolution of the perturbed horizon geometry, and are equivalent to the incompressible NS equations (\ref{NS}) \cite{Bhattacharyya:2008kq,Eling:2009pb}.
Note that since the Einstein equations are relativistic, this amounts to taking a particular non-relativistic scaling limit of the fully relativistic fluid-gravity equations \cite{Bhattacharyya:2008jc,Eling:2009sj}.
For example, one of the equations enforces the vanishing of the fractional rate of change of horizon area at lowest order
and reduces to the incompressibility condition $\partial_i v_i = 0$.

In the gravitational framework, every fluid configuration that solves the incompressible NS equations corresponds to a
particular dual bulk solution and corresponding horizon hypersurface geometry. To describe forced turbulence holographically, one can introduce a background matter term $T_{AB}$ into the Einstein field equations (\ref{Einstein}) \cite{Bhattacharyya:2008ji}. This can be used to model the large scale random force pumping energy into the fluid. The energy injected by the pumping ultimately falls into the black hole, where it is dissipated as heat. In the turbulent inertial range, the pressure and velocity are stochastic fields, implying that the dual horizon geometry and measure itself are also random.
Therefore, the statistical properties of the random horizon hypersurface (characterized by a sum over surfaces) may encode the universal statistical structure of turbulence. At the horizon, the energy dissipation is related to the flux of matter across the surface, $\int_{\Sigma} T_{AB} n^A n^B d\Sigma$ \cite{Adams:2012pj}.
Since the horizon itself is random, we were inspired to introduce the random measure in order to quantify the intermittency effects.

\section{Discussion}

We proposed an analytical formula for the scaling exponents of inertial range incompressible fluid turbulence
in any number of space dimensions $d\geq 2$.
The idea is that intermittency can be taken into account by a novel gravitational dressing of
the scale invariant Kolmogorov spectrum. Mathematically, the coupling to a random geometry with a random measure
based on a log correlated field, maps the fractal structure of the scaling exponents to a multifractal one.

There is one parameter that depends on the number of dimensions that we denoted by $\gamma(d)$.
It can be deduced knowing one moment, for instance the energy spectrum. With this knowledge
one can make infinite number of predictions.

Our formula passes the standard analytical consistency checks, such as the convexity inequality and the absence of a supersonic mode.
Its predictions agree with experimental and numerical data in two, three and four space dimensions.
The main challenge is to identify the origin of the random geometry in a more precise way and determine analytically the function $\gamma(d)$. We expect the holographic framework that inspired our formula - the relation between fluid dynamics and black hole horizon geometry - to provide a clean calculational scheme.
One may also try to calculate $\gamma$ using some physical models for the anomalous scaling, such as contributions from vortex filaments \cite{She:1994zz}, or statistical conservation laws \cite{passive}.

While equilibrium statistics is characterized by the Gibbs measure, there is yet no analog of this
for non-equilibrium steady state statistics. We speculate that  there is a general principle that allows us to consider the steady state statistics of out of equilibrium systems as a gravitationally dressed scale invariant one. If correct, this will shed much light on out of equilibrium dynamics.

Finally, it will be interesting to use the gravitational dressing to study the intermittency
effects on the anomalous scaling of the transverse structure functions and
multipoint correlation functions. Also, our proposed formula is valid for the inertial range of scales, and most likely does not incorporate statistical signatures of the dissipation range of scales. It is of interest to know whether the latter can be parametrized by a random geometry, since after all the Reynolds number is finite in nature.

\section*{Acknowledgements}
We would like to thank G. Falkovich for valuable comments.
The research of C.E. was supported by the European Research Council
under the European Union's Seventh Framework Programme (ERC Grant
agreement 307955). The work of Y.O. is supported in part by the I-CORE program of Planning and Budgeting Committee (grant number 1937/12), the US-Israel Binational Science Foundation, GIF and the ISF Center of Excellence.

\end{document}